\documentclass[prd,a4paper,preprint,preprintnumbers,superscriptaddress,nofootinbib]{revtex4-2}

\usepackage[a4paper, hdivide={1.91cm,,1.165cm}, vdivide={1.83cm,,3.0cm}]{geometry}

\usepackage[hyperfootnotes=false]{hyperref}
\hypersetup{linkcolor=red,
citecolor=green,
filecolor=cyan,
urlcolor=magenta}

\usepackage{amsfonts}

\usepackage{amsmath}
\usepackage{amssymb}

\usepackage{bbm}

\usepackage{bbold}

\usepackage{adjustbox}
\usepackage{booktabs}

\usepackage{colortbl}

\usepackage{cancel}

\usepackage{color}
\usepackage{xcolor}

\usepackage{graphicx}
\usepackage{xspace}

\usepackage{units}

\usepackage{slashed}

\usepackage{tensor}

\usepackage{gensymb}

\usepackage{cleveref}

\usepackage{tabularx}
\usepackage{multirow}

\usepackage{setspace}
\usepackage{orcidlink}

\usepackage{caption}
\usepackage{subcaption}

\usepackage{enumerate}
\usepackage{enumitem}

\usepackage[version=3]{mhchem}

\usepackage{lipsum}

\newcommand{\GG}{\langle g_s^2 G^2 \rangle}

\newcommand{\dd}{\ensuremath{\text{d}}}

\newcommand{{\ia} }{{\i}}
\newcommand{{\Ia} }{{\.I}}

\def\s2tw{{\rm sin ^2 \theta_{W}}}

\def\beq{\begin{equation}}
\def\eeq{\end{equation}}
\def\bea{\begin{eqnarray}}
\def\eea{\end{eqnarray}}

\def\vel{\left|}
\def\ver{\right|}
\def\nnb{\nonumber}
\def\lla{\left<}
\def\rra{\right>}
\def\nnb{\nonumber}

\def\es{ &=& }
\def\ar{&+& }
\def\ek{&-& }

\def\uu{\langle \bar u u \rangle}
\def\dd{\langle \bar d d \rangle}
\def\sp{\langle \bar s s \rangle}
\def\GG{\langle g_s^2 G^2 \rangle}

\def\bea{\begin{eqnarray}}
\def\eea{\end{eqnarray}}
\def\beeq{\begin{eqnarray}}
\def\eeeq{\end{eqnarray}}

\def\vel{\left|}
\def\ver{\right|}
\def\nnb{\nonumber}

\def\lla{\left<}
\def\rra{\right>}

\def\nnb{\nonumber}

\def\ba{\begin{array}}
\def\ea{\end{array}}

\def\xis0{{\Xi^{*0}}}

\def\g5{\gamma_5}

\def\es{\!\!\! &=& \!\!\!}

\def\ar{&+& \!\!\!}
\def\ek{&-& \!\!\!}

\def\olra{\stackrel{\leftrightarrow}}
\def\ola{\stackrel{\leftarrow}}
\def\ora{\stackrel{\rightarrow}}

\begin{document}


\title{Analysis of the light $J^P = 3^-$ mesons in QCD sum rules}

\author{T.~M.~Aliev\,\orcidlink{0000-0001-8400-7370}}
\email{taliev@metu.edu.tr}
\affiliation{Department of Physics, Middle East Technical University, Ankara, 06800, Turkey}

\author{S.~Bilmis\,\orcidlink{0000-0002-0830-8873}}
\email{sbilmis@metu.edu.tr}
\affiliation{Department of Physics, Middle East Technical University, Ankara, 06800, Turkey}
\affiliation{TUBITAK ULAKBIM, Ankara, 06510, Turkey}

\author{M.~Savci\,\orcidlink{0000-0002-6221-4595}}
\email{savci@metu.edu.tr}
\affiliation{Department of Physics, Middle East Technical University, Ankara, 06800, Turkey}

\date{\today}

\begin{abstract}
In this study, we investigate the masses and decay constants of the light mesons with $J^P = 3^-$ ( $\rho_3, \omega_3, K_3, \phi_3$) within the QCD sum rules method by taking into account $SU(3)$ violation effects. We state that our predictions on masses of the considered mesons are in good agreement with experimental data within the precision of the model.
\end{abstract}
\maketitle
%
%
%
\newpage
%

\section{Introduction}
\label{intro}
The quark model demonstrated remarkable success in explaining the underlying structure of hadrons. Especially for the ground states of pseudoscalar and vector mesons, the quark model provides a successful explanation for the observed spectrum. This model also predicts the existence of various radial and orbital excitations of hadrons; however, many of these still need to be confirmed through experiments. One of the main directions of the numerous experimental collaborations is the comprehensive investigation of the features of well-known light mesons as well as the search for new meson states.

A meson state with $J^P= 3^-$ means that the state has $L=2$ hence, belongs to the 1D family. These states have been observed in the light meson sectors~\cite{Baldi:1976ua,Brandenburg:1975ft,Wagner:1974gw,Aston:1988rf}, and intensive experimental studies, especially for the heavy tensor mesons, have been performed in ongoing experiments such as COMPASS \cite{Ketzer:2019wmd}, LHCb \cite{LHCb:2018roe}, BESIII \cite{BESIII:2020nme,Mezzadri:2015lrw,Marcello:2016gcn}, GlueX \cite{Austregesilo:2018mno}, and PANDA \cite{Fioravanti:2012px} collaborations. More detailed information on the current status of these states can be found in
~\cite{Ketzer:2019wmd,Padmanath:2018zqw}.  

The study of the spectroscopic parameters, like mass and decay constants of the hadrons is important to understand the dynamics of the strong interaction. Since perturbative expansions are not applicable at low energy for hadrons, phenomenological models are needed to predict meson spectroscopy. Among these models, the QCD sum rule method has been quite successful in predicting the hadron spectrum~\cite{Shifman:1978bx,Shifman:1978by}.

The comparison of the several models' predictions on the spectroscopic parameters with the experimental data allows us to test our knowledge of these states as well as understand the dynamics of the QCD in the nonperturbative domain.

In the present work, we study the mass and decay constants of the $J^{PC} = 3^{--}$ tensor mesons, such as $\rho_3(1690)$, $\omega_3(1670)$, $ K_3^\ast(1780)$, and $\phi_3(1850)$ in the framework of the QCD sum rules.

The paper is organized as follows. In Sec.~\ref{sec:2}, we derive the sum rules for the mass and decay constants of the $\bar{q} q$ nonet mesons with quantum numbers  $J^{PC} = 3^{--}$. Sec.~\ref{sec:3} is devoted to the numerical analysis of the mass sum rules for $J^{PC} = 3^{--}$  tensor mesons. The final section contains our conclusion. 

%
\section{Sum rules for the $J^{PC} = 3^{--}$ mesons}
\label{sec:2}
In this section, we derive the formulas to determine the the mass and decay constants of nonet
mesons with quantum numbers $J^{PC} = 3^{--}$ by using QCD sum rules. In this regard, we
introduce the following two-point correlation function,
\bea
\label{eq1}
\Pi_{\mu\nu\rho\alpha\beta\sigma} (p) = \left.  i \int d^4x\, e^{ip(x-y)} \lla 0 \vel
{\mbox T} \Big\{ J_{\mu\nu\rho} (x) J_{\alpha\beta\sigma}^\dag (y) \Big\}
\ver 0 \rra \ver_{y=0}~,
\eea
where $J_{\mu\nu\rho}$ is the interpolating current for the $J^{PC} = 3^{--}$ light
mesons. The current which produce these mesons from the vacuum
can be written in its simplest form as follows:
\bea
\label{eq2}
J_{\mu\nu\rho} \es {1\over 6} \bar{q} \Gamma_{\mu\nu\rho} q 
\eea
where
\begin{equation}
  \label{eq:1}
  \Gamma_{\mu\nu\rho} =  \Big[\gamma_\mu \Big(  \olra{\cal D}_{\nu} \olra{\cal D}_{\rho} +
\olra{\cal D}_{\rho} \olra{\cal D}_{\nu} \Big) +
\gamma_\nu \Big(  \olra{\cal D}_{\mu} \olra{\cal D}_{\rho} +
\olra{\cal D}_{\rho} \olra{\cal D}_{\mu} \Big) +
\gamma_\rho \Big(  \olra{\cal D}_{\mu} \olra{\cal D}_{\nu} +
\olra{\cal D}_{\nu} \olra{\cal D}_{\mu} \Big) \Big] ~,
\end{equation}
in which,
\bea
\label{eq3}
\olra{\cal D}_{\nu} \es {1\over 2} \left( \ora{\cal D}_{\nu} - \ola{\cal D}_{\nu}\right)~, \nnb \\
\ora{\cal D}_{\nu} \es \ora{\partial}_\nu - {i\over 2} g A_\nu^a t^a~,\nnb \\
\ola{\cal D}_{\nu} \es \ola{\partial}_\nu + {i\over 2} g A_\nu^a t^a~,
\eea 
and $t^a$ are the Gell-Mann matrices.

The quark content of the mesons studied in this work is as follows:
\begin{equation*}
    \begin{aligned}
     \bar{q} q =
        \begin{cases}
{1\over \sqrt{2}} (\bar{u}u - \bar{d}d) & {\rm for~}\rho_3^0\\
                  \bar{u} d             &  {\rm for~}\rho_3^+\\
          {1\over \sqrt{2}} (\bar{u}u + \bar{d}d) & {\rm for~}\omega_3^0\\           
                  \bar{s} u             &  {\rm for~}K_3^+\\
                  \bar{d} s             &  {\rm for~}K_3^0\\
                  \bar{s} s             &  {\rm for~}\phi_3^0\\
        \end{cases}
    \end{aligned}
\end{equation*}
According to sum rules method, the correlation function is calculated both in terms of hadrons (so called phenomenological part) and in terms of quark gluon degrees of freedom (theoretical part). Then, these two representations are matched and the sum rules for the relevant physical quantities are derived.

Let us start with obtaining the correlation function from phenomenological side. For this purpose, we insert a complete set of intermediate hadronic states carrying the same quantum numbers as the interpolating current, $J_{\mu\nu\rho}$, into the
correlation function. However, one needs to be careful for obtaining the phenomenological part, since the interpolating current
couples not only with the $J^P=3^-$ states, but also with the $J^P=2^+,1^-, 0^+$ states. Hence, the contributions of the unwanted states (other than $J^{P}=3^-$) should be eliminated. 
These matrix elements are defined as,
\bea
\label{eq4}
\lla 0 \vel J_{\mu\nu\rho} \ver 3^- (p ) \rra \es f_3 m_3^4
\varepsilon_{\mu\nu\rho}^{\lambda} (p) ~, \nnb \\
\lla 0 \vel J_{\mu\nu\rho} \ver 2^+ (p) \rra \es f_2 m_2^2
\Big[ p_\mu \varepsilon_{\nu\rho}^\lambda (p ) +
p_\nu \varepsilon_{\mu\rho}^\lambda (p) +
p_\rho \varepsilon_{\mu\nu}^\lambda (p) \Big] \nnb \\
\lla 0 \vel J_{\mu\nu\rho} \ver 1^- (p) \rra \es f_1 m_1
\Big[ p_\mu p_\nu \varepsilon_\rho^\lambda (p ) +
p_\nu p_\rho \varepsilon_\mu^\lambda (p) +
p_\rho p_\mu \varepsilon_\nu^\lambda (p) \Big] \nnb \\
\lla 0 \vel J_{\mu\nu\rho} \ver 0^+ (p) \rra \es f_0 \Big[
p_\mu p_\nu p_\rho \Big]~,
\eea
where $f_3$, $f_2$, $f_1$ and $f_0$ are the decay constants, $m_3$, $m_2$,$m_1$ are the masses and $\varepsilon_{\mu\nu\rho}^{\lambda}(p)$, $\varepsilon_{\mu\nu}^{\lambda}(p)$ and $\varepsilon_\mu^{\lambda}(p)$  are the polarization tensors of the corresponding mesons.
Inserting the intermediate states, and isolating the ground state
contributions from the $J^P=3^-$ states from~Eq.\ref{eq1} we get,
\bea
\label{eq5}
\Pi_{\mu\nu\rho\alpha\beta\sigma} (p) = {f_3^2 m_3^8 \over m_3^2-p^2} \sum_{\lambda=-3}^{\lambda=3} 
\varepsilon_{\mu\nu\rho}^{\lambda}(p) 
\varepsilon_{\alpha\beta\sigma}^{\lambda}(p) + \cdots
\eea
where $\cdots$ describes the contributions from $2^+$, $1^-$ and $0^+$
states. It follows from above equation that to obtain the phenomenological
part, we need to perform summations over polarizations of the corresponding
mesons, which is performed with the help of the following expressions
\cite{Jafarzade:2021vhh},
\bea
\label{eq6}
{\cal T}_{\mu\nu\rho\alpha\beta\sigma} \es \sum_{\lambda=-3}^3
\varepsilon_{\mu\nu\rho}^{\lambda} \varepsilon_{\alpha\beta\sigma}^{\lambda} \nnb \\
\es  {1\over 15}
\Bigg[ T_{\mu \nu} (T_{\beta \sigma} T_{\rho \alpha} + T_{\alpha \sigma} T_{\rho \beta} +
      T_{\alpha \beta} T_{\rho \sigma}) +
      T_{\mu \rho} (T_{\beta \sigma} T_{\nu \alpha} + T_{\alpha\sigma} T_{\nu \beta} +
      T_{\alpha \beta} T_{\nu \sigma}) \nnb \\
\ar
      T_{\nu \rho} (T_{\beta \sigma} T_{\mu \alpha} + T_{\alpha \sigma} T_{\mu \beta} +
      T_{\alpha \beta} T_{\mu \sigma})\Bigg] \nnb \\
\ek {1\over 6}
\Bigg[ T_{\mu \alpha} (T_{\nu \sigma} T_{\rho \beta}  + T_{\nu \beta}  T_{\rho \sigma}) +
      T_{\mu \beta} (T_{\nu \sigma}  T_{\rho \alpha} + T_{\nu \alpha} T_{\rho \sigma}) +
      T_{\mu \sigma} (T_{\nu \beta}  T_{\rho \alpha} + T_{\nu \alpha} T_{\rho \beta}) \Bigg]~, \nnb \\
{\cal T}_{\mu\nu\alpha\beta} \es \sum_{\lambda=-3}^3
\varepsilon_{\mu\nu}^{\lambda} \varepsilon_{\alpha\beta}^{\lambda}
= {1\over 2} \Bigg[T_{\mu \alpha} T_{\nu \beta}  + T_{\mu \beta} T_{\nu \alpha}
- {1\over 3} T_{\mu \nu} T_{\alpha \beta}\Bigg]~, \nnb \\
{\cal T}_{\mu\nu} \es \sum_{\lambda=-3}^3
\varepsilon_\mu^{\lambda} \varepsilon_\nu^{\lambda} = T_{\mu \nu}~,
\eea
where
\bea
\label{eq7}
T_{\mu\nu} = -g_{\mu\nu} + {p_\mu p_\nu \over p^2}~.
\eea
From Eqs.~\eqref{eq1} and \eqref{eq6}, it follows that the correlation function contains many structures, which can be written in
terms of numerous invariant functions $\Pi_i (p^2)$ in the following way,
\bea
\label{eq11}
\Pi_{\mu\nu\rho\alpha\beta\sigma} \es \Pi_3(p^2) {\cal T}_{\mu\nu\rho\alpha\beta\sigma} +
\Pi_2(p^2) p_\mu p_\alpha {\cal T}_{\nu\rho\beta\sigma} +
 \Pi_1(p^2)  {\cal T}_{\mu\alpha} p_\nu p_\rho p_\beta p_\sigma  +
 \Pi_0(p^2) p_\mu p_\nu p_\rho p_\alpha p_\beta p_\sigma \nnb \\
 \ar \text{ all possible permutations}~,
\eea
where subscripts in $\Pi_i$ describe the contributions of the 
$J^P = 3^-,~2^+,~1^-$ and $0^+$ mesons, respectively.
We need to isolate the contributions of $J^P=3^-$, i.e., $\Pi_3(p^2)$. For this purpose, the projection operator  ${\cal T}_{\mu\nu\rho\alpha\beta\sigma} $ is applied to the both side of Eq.\eqref{eq11}. After this operation, we get
\bea
\label{eq13}
\Pi_3(p^2) = {1\over 7} {\cal T}^{\mu\nu\rho\alpha\beta\sigma}
\Pi_{\mu\nu\rho\alpha\beta\sigma}~.
\eea
Separating the coefficient of $T_{\mu\nu\rho\alpha\beta\sigma}$ from both
representation of the correlation function we get the sum rules for the mass
and decay constant of $J^P=3^-$ tensor meson,
\bea
\label{eq14}
{f_3^2 m_3^8 \over m_3^2-p^2} = \Pi_3(p^2)~.
\eea

Having the expression of the correlation function from phenomenological part, now let us turn our attention to the calculation of it from QCD side. For this aim, we use the operator product expansion (OPE). After applying Wick theorem to the Eq.~\eqref{eq1} we get,
\bea
\label{eq8}
\Pi_{\mu\nu\rho\alpha\beta\sigma} = \left.i \int d^4x e^{ip(x-y)}
\, {\mbox Tr} \Big\{\Gamma_{\mu\nu\rho} (x) S^{ab}(x-y) \Gamma_{\alpha\beta\sigma}
(y) S^{ba}(y-x) \Big\}  \ver_{y=0}
\eea 
where $S^{ab}(x-y)$ is the light quark propagator in the coordinate space~\cite{Yang:1993bp, Wang:2008vg},
\bea
\label{eq9}
iS_q^{ab}(x)
\es \frac{i \delta^{ab}}{2\pi^2 x^4}\not\!x
  -\frac{\delta^{ab}}{12}\langle\bar{q}q\rangle
  - \frac{i \delta^{ab}g_s^2 x^2\not\!x}{2^5\times 3^5}\langle\bar{q}q\rangle^2 
 + \frac{i}{32\pi^2} g_s G_{\mu\nu}^{ab}
   \frac{\sigma^{\mu\nu}\not\!x+\not\!x\sigma^{\mu\nu}}{x^2} \nnb \\
\ar \frac{\delta^{ab}x^2}{192}\langle g_s\bar{q}\sigma Gq\rangle 
- \frac{\delta^{ab} x^4}{2^{10}\times 3^3}\langle\bar{q}q\rangle\langle g_s^2 G^2\rangle  
- \frac{m_q\delta^{ab}}{4\pi^2 x^2}
+ \frac{i m_q\delta^{ab}\not\!x}{48}\langle\bar{q}q\rangle \nnb \\
\ek  \frac{m_q \delta^{ab} g_s^2 x^4}{2^7\times 3^5}\langle\bar{q}q\rangle^2
+ \frac{m_q}{32\pi^2} g_s G_{\mu\nu}^{ab}\sigma^{\mu\nu}\ln(-x^2) 
- \frac{i m_q\delta^{ab}x^2\not\!x}{2^7\times 3^2}\langle g_s\bar{q}\sigma Gq\rangle \nnb \\ 
\ek \frac{m_q\delta^{ab}}{2^9\times 3\pi^2} x^2\ln(-x^2) \langle g_s^2 G^2\rangle
  +\cdots,
\eea
where $G_{\mu\nu}^{ab} = G_{\mu\nu}^{n} (\frac{t^n}{2})^{ab}$ is the gluon field strength tensor. 
In further calculations, we use the Fock-Schwinger gauge, i.e., $A_\mu x^\mu
= 0$. The advantage of this gauge is that the gluon field is expressed in
terms of the gluon field strength tensor as follows,
\bea
\label{eq10}
A_\mu^a (x) = {1\over 2} x_\rho G_{\rho\mu}^a(0) + {1\over 3} x_\rho
x_\sigma {\cal D}_\rho G_{\sigma\mu}^a (0) + \cdots
\eea
Putting Eqs.~\eqref{eq:1}, \eqref{eq9}, and \eqref{eq10} into Eq.~\eqref{eq8} and applying the same projection operator as used in the phenomenological side, after lengthy calculations, we obtain the QCD side of the correlation function.

In order to suppress the higher states and continuum contribution, it is necessary to perform Borel transformation over $(-p^2)$ variable from both sides of correlation function. Finally, matching the results of correlation function obtained from QCD and hadron side, we get the mass sum rules for corresponding tensor meson, $J^P = 3^-$ as,
\begin{equation}
  \label{eq:2}
  f_3^2 m_3^8 e^{-m_3^2/M^2} = \Pi_3^{(B)}
\end{equation}
where $\Pi_3^{B}$ is the Borel transformed form of the invariant function, which is,
\bea
\label{eq15}
\Pi_3^{B}
\es
 - {5 \over (2^4\times 3^2) \pi^2}
 e^{ - s_0/M^2} \GG M^2 m_{q_1} m_{q_2} \nnb \\
\ek {1\over (2^4 \times 3^3 \times 5 \times 7)\pi^2} M^4
\Bigg[ 91 \GG \Big(\Gamma[2,0, s_0/M^2] \Big)
- 36 M^4 \Big(\Gamma[4,0, s_0/M^2] \Big) \Bigg] \nnb \\
\ar {1\over (2^8 \times 3^2) M^2 \pi^2}
\Bigg\{ 5 \GG \Big( \GG + 16 M^4 \Big) + 48 M^8 \Big( \Gamma[3, 0,s_0/M^2] \Big)
\Bigg\} m_{q_1} m_{q_2} \nnb \\
\ek {5 \over (2^9 \times 3^4) M^4})
\GG \Big( 5 \GG + 192 M^4 \Big)
\Big( m_{q_1} \langle \bar q_2 q_2 \rangle +
       m_{q_2} \langle \bar q_1 q_1 \rangle \Big) \nnb \\
\ek
    {5 \over (2^6 \times 3^6) M^4}  g_s^2 \Big(5 \GG + 192 M^4 \Big)
\Big( \langle \bar q_1 q_1 \rangle^2 + \langle \bar q_2 q_2
\rangle^2 \Big) m_{q_1} m_{q_2}~.
\eea
Here $M^2$ is the Borel-mass parameter, $s_0$ is the continuum threshold, $m_q$ is the mass of the light quark, and  
\bea
\Gamma[n,0,s_0/M^2] = \int_0^{s_0/M^2} dt t^{n-1} e^{-t}~,\nnb
\eea
is the generalized incomplete gamma function.

Differentiating both sides of Eq. (\ref{eq15}) with respect to $-{1\over M^2}$ and
dividing it by itself, we get the QCD sum rules for the mass of the $J^P=3^-$ tensor mesons, 
\bea
\label{eq16}
m_3^2 \es {1\over \Pi_3^{B}} {d\Pi_3^{B} \over d(-1/M^2)}~. \nnb 
\eea

Once the mass is determined, we can use it as an input parameter and obtain the decay constant $f_3$ of tensor mesons using
Eq. (\ref{eq15}).

In addition, to determine the spectroscopic parameters of $J^P=3^-$ tensor mesons
we also used another approach, which is based on the corporation of
the sum rules with the least square method. The main idea of this approach
is the minimization of the square difference of the phenomenological and OPE
parts of the correlation function as given below,
\bea
\sum_{i=1}^N {\vel f_3^2 m_3^8 e^{-m_3^2/M_i^2} - 
\Pi_3^{B}(M_i^2,s_0) \ver^2 \over N}~,\nnb
\eea
By applying two-parameter ($M^2$ and $s_0$) fitting, we try to minimize the
above expression using appropriate sets of the parameters $\{ m_3^2 \}$ and $\{ s_0 \}$. In the following discussions, we will mention this method as Method-B.
\section{Numerical analysis}
\label{sec:3}
In this section, we perform numerical analysis to determine the mass and decay constants of the $J^P=3^-$ tensor mesons using the expressions obtained in the previous section. It follows from Eq.(\ref{eq15}) that the sum rule involves input parameters such as light-quark masses, quark and gluon condensates. We use the standard values for the 
quark and gluon condensates, i.e., $\GG = 4 \pi^2 \times 0.012~GeV^4$, $\uu = \dd = -(0.24~GeV)^3$ \cite{Shifman:1978bx}, and $\sp = - 0.8 \uu$ \cite{Belyaev:1982cd}. For the mass of the strange quark, $\overline{MS}$ value is used as $\overline{m_s}(1~GeV) = 0.126~GeV$~\cite{PDG:2022pth}.

In addition to these input parameters, the sum rule contains two more auxiliary parameters, namely, the continuum threshold, $s_0$ and the Borel mass parameter, $M^2$. Obviously, physically measurable quantities like mass should be independent of these parameters. For this reason, we should determine the working regions of $M^2$ and $s_0$.

\begin{table}[th]
\renewcommand{\arraystretch}{1.2}
\setlength{\tabcolsep}{6pt}
  \begin{tabular}{lcc}
    \toprule
            &    $M^2\,(GeV^2)$ & $s_0\,(GeV^2)$  \\
\midrule
$\rho_3$    &   $1.3 \div 1.5$  & $4.0 \div 4.2$  \\
$\omega_3$  &   $1.3 \div 1.5$  & $4.0 \div 4.2$  \\
$K_3$       &   $1.4 \div 1.6$  & $4.4 \div 4.6$  \\
$\phi_3$    &   $1.5 \div 1.7$  & $4.8 \div 5.0$  \\
    \bottomrule
  \end{tabular}
\caption{The working regions for Borel mass $M^2$ and continuum threshold $s_0$.}
\label{tab:1}
\end{table}
\begin{table}[b]
\begin{adjustbox}{center}
\renewcommand{\arraystretch}{1.2}
\setlength{\tabcolsep}{6pt}
  \begin{tabular}{l|ccc|cc}
    \toprule
 & \multicolumn{3}{c|}{Mass in $GeV$} & \multicolumn{2}{c}{Decay Constant}\\
            & Method-A  
            & Method-B  
            & Exp. \cite{PDG:2022pth}
            & Method-A
            & Method-B \\
\midrule
$\rho_3$    &   $1.76\pm 0.01$  & $1.78 \pm 0.02$  & $(1.680 \pm 0.020)$  & $(1.17\pm 0.01) \times 10^{-2}$ & $(1.17 \pm 0.02) \times 10^{-2}$ \\
$\omega_3$  &   $1.76\pm 0.01$  & $1.78 \pm 0.02$  & $(1.667 \pm 0.004)$  & $(1.17 \pm 0.01) \times 10^{-2} $ & $(1.17 \pm 0.02)\times 10^{-2}$ \\
$K_3$       &   $1.82\pm 0.02$  & $1.83 \pm 0.02$  & $(1.779 \pm 0.008)$  & $(1.26 \pm 0.01) \times 10^{-2}$ & $(1.25 \pm 0.01) \times 10^{-2} $ \\
$\phi_3$    &   $1.86\pm 0.01$  & $1.88 \pm 0.02$  & $(1.854 \pm 0.007)$  & $(1.36 \pm 0.01) \times 10^{-2} $ & $(1.35 \pm 0.01) \times 10^{-2}$ \\
    \bottomrule
  \end{tabular}
\end{adjustbox}
\caption{Our predictions on mass and decay constants of the $J^P=3^-$ tensor
mesons. For completeness we also present the experimental values of the
mesons under consideration.}
\label{tab:2}
\end{table}
The upper bound of $M^2$ is determined by requiring the pole dominance over the higher states and continuum contribution. This is determined by
the ratio,
\bea
\label{eq17}
{\Pi_3^{B}(s_0,M^2) \over \Pi_3^{B}(\infty,M^2)}~
\eea
where $\Pi^B_3(s_0,M^2)$ is the Borel-transformed and continuum subtracted invariant function $\Pi^{OPE}$. We demand that the pole contributions constitute more than $50\%$ of the total result. The minimum value of $M^2$ is obtained by requiring that the OPE should be convergent. For this aim the following ratio is considered,
\bea
\label{eq18}
{\Pi^{B (highest~dimension)}(s_0,M^2) \over \Pi(s_0,M^2)}~,
\eea
where $\Pi^{B (\text{condensates})}(s_0,M^2)$ is the contributions of the condensate terms.
We require that the total contributions of condensates should be less than $30\%$ of the total result.

\begin{figure}[h]
  \centering
  \begin{subfigure}{\textwidth}
\includegraphics[width=0.49\textwidth]{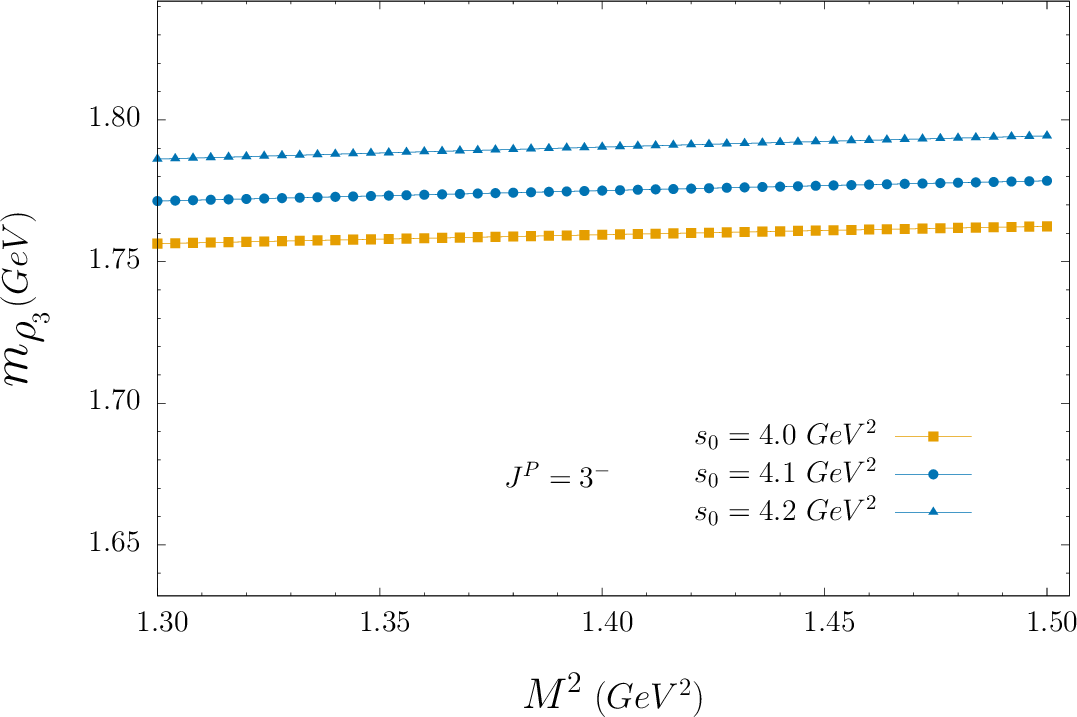}
\includegraphics[width=0.49\textwidth]{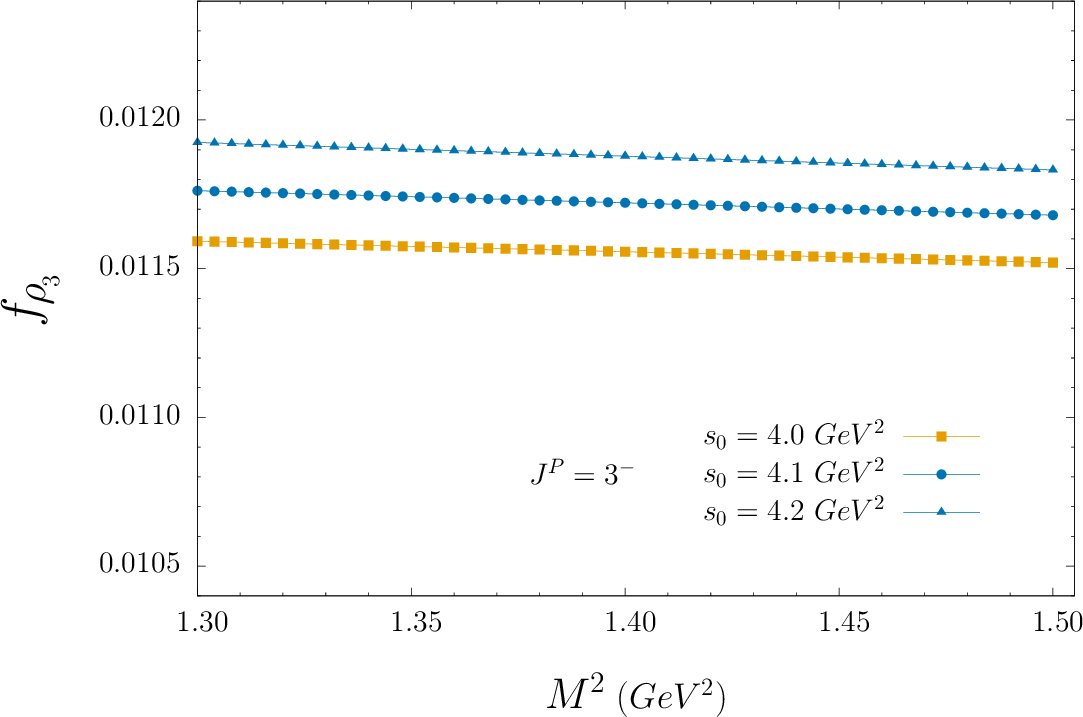} 
\caption{
The dependencies of mass and residue of the $\rho_3$ meson  on Borel mass square $M^2$ at three fixed values of the continuum threshold $s_0$.
}
\label{fig:fig1}
\end{subfigure}

\begin{subfigure}{\textwidth}
\includegraphics[width=0.49\textwidth]{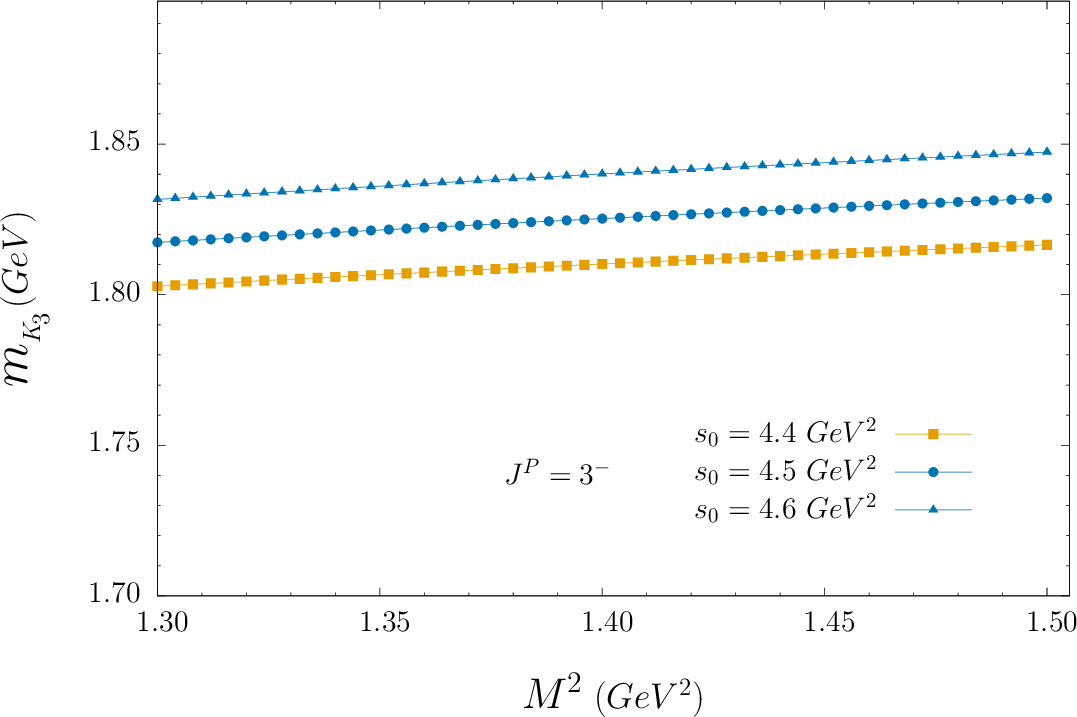}
\includegraphics[width=0.49\textwidth]{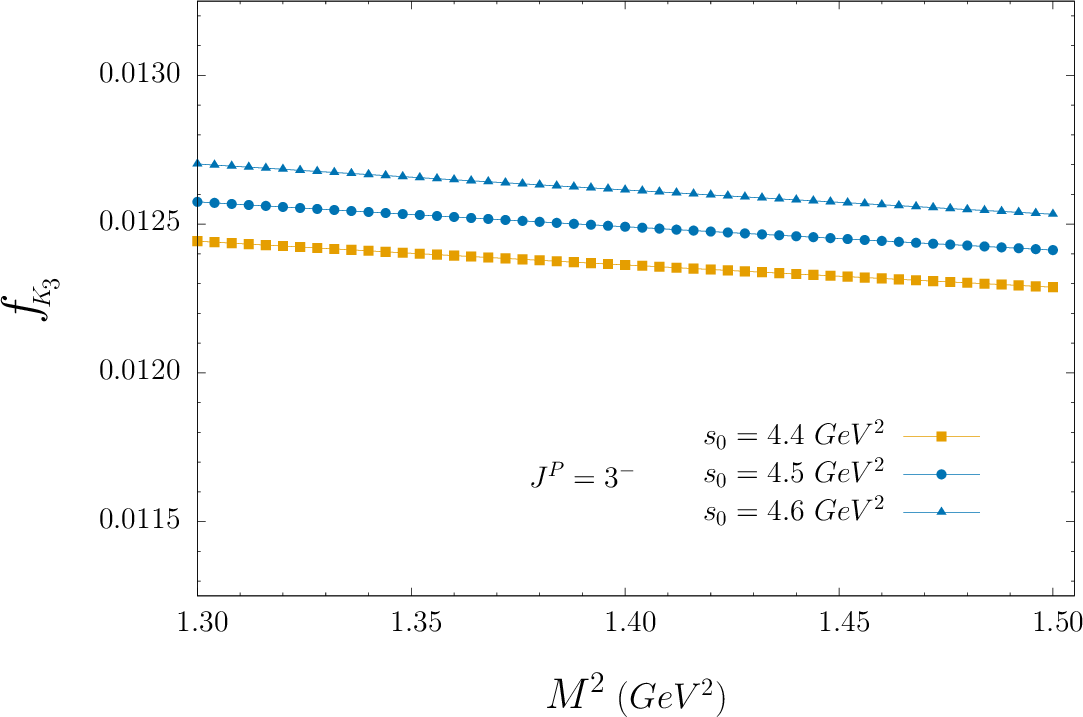} 
\caption{
The dependencies of mass and residue of the $K_3$ meson  on Borel mass square $M^2$ at three fixed values of the continuum threshold $s_0$.
}
\label{fig:fig2}
\end{subfigure}
  
\begin{subfigure}{\textwidth}
\includegraphics[width=0.49\textwidth]{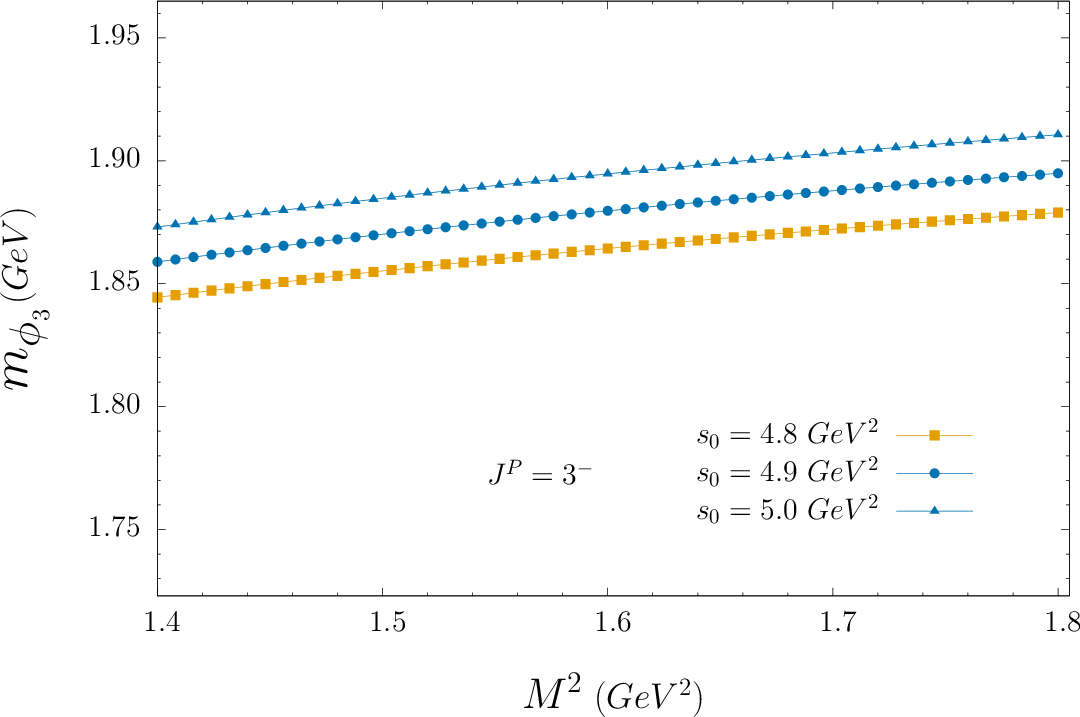}
\includegraphics[width=0.49\textwidth]{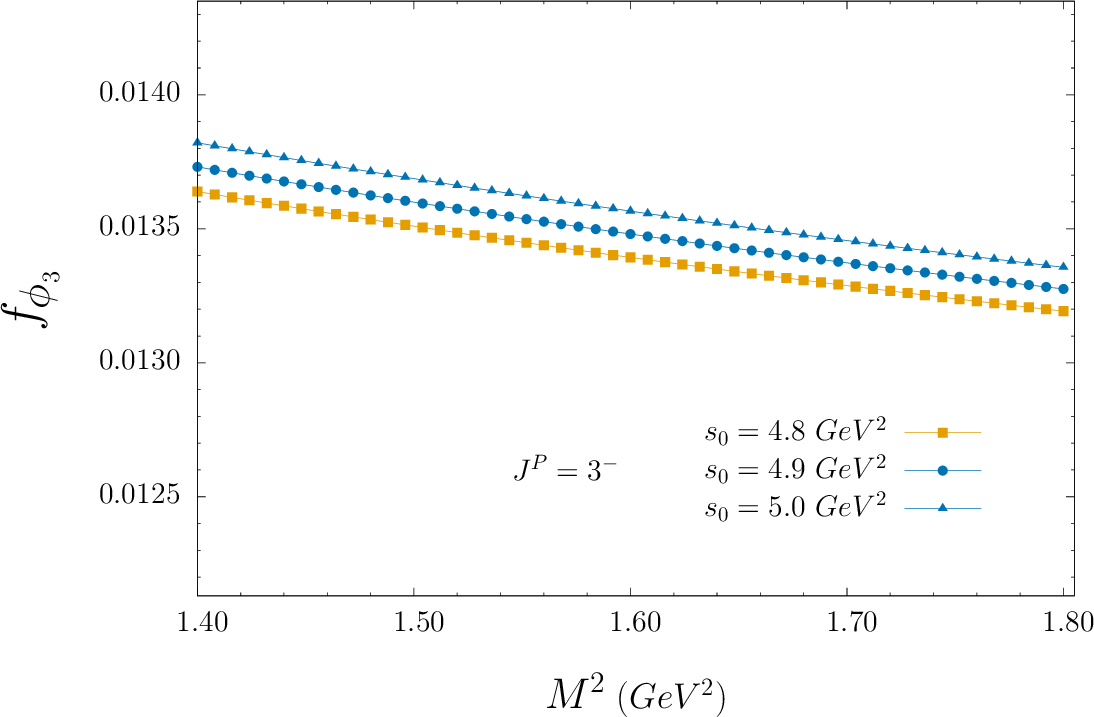} 
\caption{
The dependencies of mass and residue of the $\phi_3$ meson  on Borel mass square $M^2$ at three fixed values of the continuum threshold $s_0$.
}
\label{fig:fig3}
\end{subfigure}
\end{figure}
The continuum threshold $s_0$ is obtained by requiring that the variation of
the masses of the $J^P$ tensor meson state with respect to $M^2$ should be
minimum.  Using these conditions, we determine the working regions of $s_0$
and $M^2$ for the mesons considered. These are presented in Table~\ref{tab:1}.

To show the stabilities of the working regions of $M^2$, in Figs. \ref{fig:fig1}, \ref{fig:fig2} and \ref{fig:fig3},  we present the dependencies of the mass
and residue of $\rho_3 (\omega_3)$, $K_3$ and $\phi_3$ on $M^2$ at several fixed values of $s_0$. Examining these figures, we see that the values of mass and residues exhibits good stability when $M^2$ varies in their corresponding working regions. And, one can determine these quantities. These values are presented in Table~\ref{tab:2}. This method is denoted as Method-A in this table. Moreover, we also calculated the mass and decay constants of the $J^P=3^-$ tensor mesons with the help of the least square method (Method-B). We observe that the predictions on mass of the considered mesons of both approaches are quite close to each other.
%
%
%

%
%
In Table~\ref{tab:2}, we also present the experimental values of the mesons under consideration. When compared our findings with the experimental ones, we see that our results on mass values of the tensor mesons are quite compatible.


%
\section{Conlusion}
In conclusion, we have determined the mass and decay constants of the $J^P=3^-$ tensor mesons by considering the $SU(3)$ violation effects. Our predictions on the masses of the tensor mesons are in good agreement with the experimental data within the precision of the model. This finding verifies that the QCD sum rules method works quite succesfully in the analysis of the physical parameters of the higher $J$ states. The obtained decay constants can be used for further studies of the strong and electromagnetic decays of the $J^P=3^-$ tensor mesons.
%
%



\bibliographystyle{utcaps_mod}
\bibliography{/Users/sbilmis/LateX/mybib/all.bib}


\end{document}